\documentclass[preprint,number,paper,12pt]{elsarticle}

\usepackage{simplemargins}

\setleftmargin{2cm}
\setrightmargin{2cm}
\settopmargin{2cm}
\setbottommargin{2cm}

\usepackage{subfigure}
\usepackage{epsfig,amsfonts}
\usepackage{lscape}
\usepackage{amssymb}
\usepackage{subfigure}
\usepackage{epsfig,amsfonts}
\usepackage{lscape}
\usepackage{capt-of}
\usepackage{graphics}
\usepackage{graphicx}
\usepackage{epsfig}
\usepackage{lscape}
\usepackage{multirow}
\usepackage{amsthm}
\usepackage{amsmath}
\usepackage{mathrsfs}
\usepackage{bm}

\usepackage{capt-of}
 \usepackage{graphics}
 \usepackage{graphicx}
 \usepackage{epsfig}
\usepackage{lscape}
\usepackage{multirow}

\usepackage{amssymb}
 \usepackage{amsthm}
 
\usepackage{amsmath}
\usepackage{ mathrsfs }
\usepackage{lineno}

\usepackage{anyfontsize}
\usepackage{helvet}

\biboptions{compress, square}

\journal{...}

\begin{document}

\begin{frontmatter}



\title{Experimental observation of a large, low frequency Lamb band gap induced by periodic cross-like cavities in a polymer waveguide}


\author{M. Miniaci$^{a,*}$, M. Mazzotti$^{b}$, M. Radzie\'nski$^c$, N. Kherraz$^a$, P. Kudela$^c$, W. Ostachowicz$^c$, B. Morvan$^a$, F. Bosia$^{d,*}$, N. M. Pugno$^{e,f,g}$}

\address{$a$) University of Le Havre, Laboratoire Ondes et Milieux Complexes, UMR CNRS 6294, 75 Rue Bellot, 76600 Le Havre, France}

\address{$b$) Civil, Architectural \& Environmental Engineering (CAEE) Department, Drexel University, 3141 Chestnut St., Philadelphia, PA 19104, USA}

\address{$c$) Institute of Fluid-Flow Machinery, Polish Academy of Science, Fiszera 14 st. 80-231 Gda\'nsk, Poland}

\address{$d$) Department of Physics and Nanostructured Interfaces and Surfaces Centre, University of Torino, Via Pietro Giuria 1, 10125 Torino, Italy}

\address{$e$) Laboratory of Bio-Inspired and Graphene Nanomechanics, Department of Civil, Environmental and Mechanical Engineering, University of Trento, Via Mesiano 77, 38123 Trento, Italy}

\address{$f$) School of Engineering and Materials Science, Queen Mary University of London, Mile End Road, London E1 4NS, United Kingdom}

\address{$g$) Ket Lab, Edoardo Amaldi Foudation, Italian Space Agency, Via del Politecnico snc, 00133 Rome, Italy}

\address{$*$ Correspondence to: marco.miniaci@gmail.com, fbosia@unito.it}


\begin{abstract}
The quest for large and low frequency band gaps is one of the principal objectives pursued in a number of engineering applications, ranging from noise absorption to vibration control, to seismic wave abatement. For this purpose, a plethora of complex architectures (including multi-phase materials) and multi-physics approaches have been proposed in the past, often involving difficulties in their practical realization.

To address this issue, in this work we propose an easy-to-manufacture design able to open large, low frequency complete Lamb band gaps exploiting a suitable arrangement of masses and stiffnesses produced by cavities in a monolithic material. The performance of the designed structure is evaluated by numerical simulations and confirmed by Scanning Laser Doppler Vibrometer (SLDV) measurements on an isotropic polyvinyl chloride plate in which a square ring region of cross-like cavities is fabricated. The full wave field reconstruction clearly confirms the ability of even a limited number of unit cell rows of the proposed design to efficiently attenuate Lamb waves. In addition, numerical simulations show that the structure allows to shift of the central frequency of the BG through geometrical modifications. The design may be of interest for applications in which large BGs at low frequencies are required.

\end{abstract}

\end{frontmatter}


\section{Introduction}
One of the main problems facing physicist and engineers working in the field of metamaterials is to achieve vibration damping and control over large, low frequency ranges. This is true in fields ranging from noise absorption \cite{Jimenez_APL_2016, Li_Assouar_APL_2016, Morandi2016294, AureganJASA2016} to seismic wave abatement \cite{Colombi_SR_2016, Miniaci_NJP_2016}. Phononic crystals (PCs) and acoustic metamaterials (AMMs), generally made of periodically distributed inclusions in a matrix (or hosting material), are of particular interest because of their ability to act as stop-band filters, i.e. attenuate mechanical waves over entire frequency bands, commonly known as band gaps (BGs) \cite{Hussein_AMR_2014, Pennec2010229, craster2012acoustic}.

Bragg scattering and local resonance are the mechanisms mainly exploited for BG nucleation. BGs due to Bragg scattering arise from the wave diffraction by periodic inclusions and thus occur at wavelengths of the order of the unit cell size, whilst local resonance is related to the vibration of individual elements within the medium, and is thus independent of the spatial periodicity of the lattice \cite{Baravelli20136562}. In some cases, Bragg BGs can be coupled with hybridization BGs \cite{Croenne_AIP_2011, Kaina_SR_2013}, due to resonating "inhomogeneities" (Fano interference effects \cite{Kosevich_PRB_Fano}, electrical resonances \cite{Bergamini_JAP_2015, kherraz2016controlling}, etc.).

The concept of a BG naturally lends itself to applications involving vibration damping, acoustic filtering and wave attenuation, as well as waveguiding \cite{Pennec2010229, Khelif2005}. Applications at high frequencies include wave filters, couplers, sensing devices, wave splitters, demultiplexers, etc. \cite{pennec2004tunable, sukhovich2009experimental}. A nonexhaustive list of low-frequency applications includes acoustic absorption, vibration shielding, subwavelength imaging, cloaking, etc \cite{craster2012acoustic, Deymier_Book}.

With these applications in mind, a plethora of different architectures (including multi-phase materials) and multi-physics approaches have been proposed in recent years \cite{Pennec2010229, Wang20132019, Andreassen2015187, Bavencoffe_Attenuation}, but many involve considerable practical difficulties in their realization due to their inherent complexity.

Recently, particular attention has been focused on phononic plates because of their potential technological applications, ranging from microelectromechanical systems to nondestructive evaluation \cite{Miniaci_PhysRevLett, Gliozzi_APL, TT_Wu_Phononic_plate_waves, Hsu_Propagation_of_Lamb, Mohammadi2011524, Celli2014114, PhysRevLett_105_074301}. Among others, phononic plates made of periodic distributions of studs or gratings on the surface \cite{Casadei_Piezo, Yu201312, PhysRevB_79_104306} as well as those realized by a periodic distribution of empty cavities perpendicular or parallel to the propagation plane \cite{TT_Wu_Phononic_plate_waves, Hsu_Propagation_of_Lamb, Chen2012920, Wang20132019} have been studied.

In addition to theoretical predictions, experimental measurements of BGs in phononic plates have been performed. In this context, Brunet et al. \cite{Brunet_JAP_2008} reported BGs for Lamb waves propagating in rectangular and square arrays of holes drilled in a silicon phononic plate, whilst Bonello et al. \cite{Bonello_Lamb_waves_in_plates_covered_by_a_two_dimensional_phononic_film} proved the existence of BGs for Lamb waves in silicon plates coated by a very thin phononic film. Publications dealing with BGs for Lamb waves generated in a plate with a periodic grating on the surface have also appeared \cite{Bavencoffe_Attenuation, Bavencoffe2013313} with the aim of quantitatively verifying the relation between the width of the BG and the depth of the grooves. Finally, several studies have considered and shown that filtering and waveguiding properties are achievable in plates with stubs, including bi-phase materials \cite{TT_Wu_Phononic_plate_waves, Hsu_Propagation_of_Lamb, Casadei_Piezo, PhysRevB_79_104306, Oudich_PhysRevB_84_165136, Hsiao_PRE_2007}.

In these approaches, geometrical/physical complexity is often inevitable, leading to considerable complications in their practical realization, which requires unit cells capable of providing the desired dynamic performance, with realistic structural features, such as being compact, lightweight and easy to manufacture. In this work, we propose phononic plate that is simple to fabricate and is capable of opening large, low frequency complete Lamb BGs, exploiting a suitable arrangement of masses and stiffnesses produced by cavities in a monolithic material. This simple but efficient design capable of producing large and low frequency Lamb band gaps when cross-like holes are considered, showing that a sort of "mass centrifugation" process is beneficial. The idea is supported by numerical calculations and experimental Scanning Laser Doppler Vibrometer (SLDV) measurements on an isotropic PolyVinyl Chloride (PVC) plate with a square ring region of cross-like holes. The full wave field reconstruction existing below, within and above the BG frequencies clearly confirms the ability of the proposed design to stop Lamb waves in the desired frequency range.

The paper is organized as follows: first we present extensive numerical simulations aimed at finding suitable plate configurations capable of nucleating large BGs within the optimal operative frequency range of the SLDV using Finite Element simulations (FEM).

Through-the-thickness holes with different geometries are considered as scatterers (inclusions). Next, finite element time-transient analysis is performed to study the influence of the phononic region length (number of unit cells) on wave propagation. Finally, SLDV measurements on ad-hoc machined phononic polyvinyl chloride (PVC) plate are presented and compared to the numerically predicted results. The calculated BGs are verified experimentally and the direct observation of scattering phenomena below, within and above the BGs is described.

\section{Design and numerical analysis}

In this section, the description of the phononic plate design is provided. Different geometries of the scatterers, made of through-the-thickness cavities, are considered to find Bragg-type BGs in the frequency range $0 - 70$ kHz.

In view of the experimental phase, PVC has been chosen as the material matrix since it can be easily machined. The PVC mechanical properties are the density $\rho_{PVC} = 1430$ kg$/$m$^3$, the Young's modulus $E_{PVC} = 3$ GPa and the Poisson's ratio $\nu_{PVC} = 0.4$ \cite{Miniaci2014}.

A lattice parameter of $a = 20$ mm is chosen for the unit cell and plate thickness $H = 0.4 a$, so that the lowest Bragg BGs relative to the $A0$ mode are expected at around $14$ and $36$ kHz, corresponding to the half and to the full wavelength of the $A0$ propagating mode, respectively. The $A0$ mode is chosen as it is easy to excite via piezoelectric (PZT) sensors and to measure via SLDV. Once the lattice parameter has been set, band structures for different types of structures are computed. The unit cells are shown in Fig. \ref{Fig1}: relative to (a) circular, (b) square and (c) rounded cross-like cavities. Geometrical parameters (see Table \ref{GeomPar}) are chosen so as to guarantee a constant filling fraction $f_f = 41\%$ for all the three models.

The corresponding band structures are shown in Figs. \ref{Fig1}d-f along with the mode shapes for some specific branches (\ref{Fig1}g-j). The dispersion diagram is computed along the three high symmetry directions of the first irreducible Brillouin zone (Fig. \ref{FIBZ}) exploiting the Bloch-Floquet theorem by using COMSOL MultyPhysics. Band structures are reported in terms of reduced wavevector $k^*$ with wavenumber components $k^* = [k_x a/\pi; k_y a/\pi]$. A detailed description of the finite element procedure can be found in Ref. \cite{Miniaci2014}.

Phononic plates with circular and square holes do not nucleate any BG, whereas the unit cell with a rounded cross-like hole exhibits two complete BGs (highlighted as light grey regions). In order to shed light on this different behaviour we highlight the in-plane or out-of-plane polarization of the modes in the dispersion diagrams by defining a polarization factor:

\begin{equation}
p=\frac{\int_V (|u_z |)^2 dV}{\int_V (|u_x |^2+|u_y |^2+ |u_z |^2)dV}, 
\end{equation}

\noindent where $V$ is the volume of the unit cell, $u_x$, $u_y$ and $u_z$ are the displacement components along $x$, $y$ and $z$ axes, respectively. The points of the dispersion curves are shaded accordingly, with colours varying from $p = 0$ (blue) to $p = 1$ (red). Thus, colours close to red indicate vibration modes that are predominantly polarized out-of-plane, while colours close to blue are predominantly polarized in-plane.

This allows us to infer that the different arrangement of the cavities is responsible for the shift of some dispersion bands while leaving others unaltered. For instance, Figs. \ref{Fig1}d,e highlight the shift occurring for two modes: one with a mainly in-plane deformation mechanism (blue rectangle) and another one with a mainly out-of-plane deformation mechanism (red rectangle). The shift is clearly due to the equivalent stiffness variations occurring due to the cavity geometry (see Fig. \ref{Fig1}g,h). Introducing a cross-like cavity can accentuate this process and lead to the opening of two BGs (Fig. \ref{Fig1}f) with a smaller deformation of the unit cell for the same branches (Fig. \ref{Fig1}j).

Results hold true for varying plate thickness with the same type of mechanical effects. Fig. \ref{Fig3} shows a parametric analysis in which band structures for the unit cell with a cross-like cavity have been computed increasing the thicknesses from $H = 0.2 a$ to $H = 1.2 a$ in steps of $0.2 a$. Results show that the plate thickness plays a critical role for (i) the nucleation of BGs, (ii) the frequency shift of the curves, and thus of the upper and/or lower BG boundaries, as well as (iii) polarization of the unit cell modes (as for the previous case, the color scale denotes here the in-plane$/$out-of-plane contribution to the total modal displacement). If the thickness of the plate is relatively small (see Fig. \ref{Fig3}a for $H \le 0.4 a$) the modes are characterized by mainly in-plane or -out-of-plane deformation (curves are either red or blue) whereas for thicknesses $H \ge 0.8 a$) (see Fig. \ref{Fig3}b) deformation modes with non-negligible in-plane and out-of plane components are present.

The dependence of the BGs with respect to the plate thickness are summarized in Fig. \ref{WAZ_BGs} where the lower and upper bounds and width of the BGs are reported as a function of $ H/a $ (in steps of $0.1 a$). No BG appears in the system when $ H/a = 0.2$ whereas a single large BG from approximately $20.70$ kHz to $28.00$ kHz is present when $H/a = 0.7$. In all the other cases two total BGs are present. The plate thickness giving rise to the widest total BG is for $H/a = 0.6$, corresponding to a relative width of $36 \%$($f_{BG_L} \in [20.60 - 27.95]$ kHz and $f_{BG_U} \in [28.42 - 32.52]$ kHz, (highlighted in red in Fig. \ref{WAZ_BGs}). In this case, the two BGs are connected, except for an almost flat band between them.

In conclusion, results show that as the plate thickness increases, the frequency range covered by BGs increases up to the largest value at $H/a = 0.6$, which is therefore the chosen thickness for the manufacturing of experimental samples. To determine the necessary number of repeated unit cell to obtain efficient signal damping in experiments, finite element time transient analysis are performed (using ABAQUS). Numerical models with unit cells such as the one presented in Fig. \ref{Fig1}c, arranged in concentric square rings with $1, 2, 3$ and $4$ rows in a specific portion of the PVC plate, as shown in Fig. \ref{Rows_design}, are considered.

First, a wave propagation analysis is performed to evaluate the screening power of the phononic region as a function of the number of rows of unit cells over a large frequency range. Guided waves are excited by means of an imposed orthogonally displacement ($1 \times 10^{-6}$ mm in the $z$-direction) at point $A$ (see Fig. \ref{Rows_design}a). A pulse of $2$ sinusoidal cycles centred at $50$ kHz and modulated by a Hanning window (Fig. \ref{Pulses_PVC_plate}a) is chosen since it generates Lamb waves over a large frequency spectrum.

The out-of-plane displacement detected at point $R2$ for the four considered cases along with their Fourier transforms are shown in Fig \ref{TimeHistory_FFT_Fc50kHz_Nc2}. The system is modelled as linear elastic, so that the signal attenuation is totally due to the Bragg scattering mechanism. The results clearly show that waves with a frequency content falling outside the BG (gray rectangle) propagate through the phononic region reaching the monitoring point $R2$ without substantial attenuation as the number of cross-like cavities increases, whilst when the frequency content of the propagating waves falls inside the BG, the phononic crystal ring region inhibits the wave propagation so that the displacement at point $R2$ is much smaller as the number of cross-like cavities increases. Results show that the destructive interference mechanism starts taking place after two unit cells and that as the number of unit cells increases the edges of the attenuation frequency regions become sharper. The results are presented in semi-logarithmic scale and prove that each additional row of unit cells introduces an attenuation factor of about $4$, allowing to infer that a plate with four rows of unit cells can reduce incident wave amplitudes by more than one order of magnitude.

The dynamics of the plate is further investigated assigning as excitation pulse $21$ sinusoidal cycles centred at $27.5$ kHz (i.e., inside the BG) and modulated by a Hanning window (Fig. \ref{Pulses_PVC_plate}b). The full wavefield reconstruction of the Von Mises stress after $780 \mu s$ from the excitation is reported in Fig. \ref{VonMises_stress_ParametricoFori}. Green and red colours correspond to zero and maximum stresses induced in the plate. The area surrounded by four rows of cross-like cavities is practically free from stresses. For this reason, the phononic plate with four rows of cross-like cavities is chosen for the experimental phase of the study. In particular, the cavities are distributed over a square frame of width $4 \times a = 80$ mm. An homogeneous area of $120 \times 120$ mm$^2$ is enclosed by the annular phononic crystal region.

\section{Experimental measurements and full wavefield reconstruction}

The sample used for experimental analysis consists of a standard $1000 \times 500 \times 12$ mm$^3$ (PVC) plate subjected to a machining process whereby $160$ hollow rounded cross-cylinder inclusions were drilled, as shown in Fig. \ref{piastra_PVC_reale}a. The cavities cover a total area of $64$ cm$^2$ and are arranged in $4$ concentric rings which leave a homogeneous $120 \times 120$ mm$^2$ PVC region at the centre. The manufacturing process requireds a tolerance of $0.1$ mm, which slightly modifies the band structure of the system \cite{Miniaci2014}.

The adopted experimental set-up consists of a scanning laser Doppler vibrometer (SLDV) PSV 400 3D by Polytec (see Fig. \ref{piastra_PVC_reale}b) which allows waveform digitization up to $70$ kHz, which is sufficient for detecting and visualizing ultrasonic waves in the frequency range of interest since the probe beam size is small enough to detect the motion at a single point on the surface of the phononic plate \cite{Kudela2015456, ostachowicz2012guided}. Elastic guided waves are induced by means of a ceramic piezoelectric disk of diameter $10$ mm made of Sonox$^\text{{\textregistered}}$ by CeramTec$^\text{{\textregistered}}$ bonded to the surface of the investigated sample (point $A$ in Fig. \ref{piastra_PVC_reale}a) using commercial super-glue. The pulse is excited by a TGA1241 generator by Thurlby Thandar Instruments through an EPA-104 amplifier by Piezo Systems$^\text{{\textregistered}}$ Inc, inducing a $200$ V$_{\text{pp}}$ signal. In order to improve measurement accuracy the investigated specimen is covered with self-adhesive retro-reflective film by ORALITE$^\text{{\textregistered}}$ enhancing the laser vibrometer signal level in each measurement point regardless of the incidence angle of the measurement beam.

The investigated spatial grid covers the entire phononic surface (see Fig. \ref{piastra_PVC_reale}a) and consists of $500 \times 500$ equally spaced grid points. The laser vibrometer is perpendicularly positioned $1$ m away from the investigated surface and only one SLDV head is used to perform the out-of-plane measurements of the velocity amplitudes over the target area (Fig. \ref{piastra_PVC_reale}b). Multiple ($512$) measurements are performed and averaged for each node, to filter out part of the noise.

A set of verification experiments is conducted to evaluate the elastic wavefield of the phononic plate at some frequencies of interest highlighted by the numerical study. Based on this analysis, sine functions with $21$ Hanning modulated sine cycles with central frequencies of $15$, $27.5$ and $42.5$ kHz are used as input signals (Fig. \ref{Pulses_PVC_plate}b-c) to reconstruct the wavefield for excitation below, within and above the BGs, respectively.

The excitation signal is used to trigger the data acquisition process. Out-of-plane velocity time-histories are acquired starting from laser beam reflections back to the scanning head and their Doppler frequency shift detected by interferometers. Upon completion of measurements at all grid points, the recorded responses are post-processed to obtain full images of the propagating wavefield aggregating single-point measurements within the region of inspection. The knowledge of the velocity time histories at all grid points (which can be integrated to obtain displacement time histories) allows the reconstruction of the time-evolving wavefields established in the scanning domain \cite{ostachowicz2012guided}.

Figure \ref{Wavefield_Reconstruction_Cfr_Num_Exp} shows the comparison of the experimental wavefield reconstruction and the corresponding calculated ones for the three aforementioned excitations (Figs. \ref{Pulses_PVC_plate}a-c). At frequencies below the BGs (Figs. \ref{Wavefield_Reconstruction_Cfr_Num_Exp}a,b), Lamb waves travel through the periodic lattice relatively undisturbed. Limited reflection, scattering, or other losses can be observed, indicating that the phononic region does not interfere significantly with the wave motion.

When operating at a frequency within the BGs, instead, the destructive interferences due to the Bragg scattering occurring within this region becomes large. Figs. \ref{Wavefield_Reconstruction_Cfr_Num_Exp}c,d clearly show that waves are reflected in the plate portion around the phononic region, mainly between the transmitting PZT and the lower edge of the unit cell ring. This behaviour is accompanied by an extremely low transmission due to the absence of detectable wave amplitudes inside the phononic region.

Above the BG, the SLDV again registers transmission inside the phononic region, allowing the wavefield reconstruction at the same intensity scale. However, in this case unit cells scatter the wave field, resulting in an observable delay in the wave propagation (Fig. \ref{Wavefield_Reconstruction_Cfr_Num_Exp}e,f). In this case, despite the scattering, the phononic region does not cause significant attenuation of the wave field.

Thus, numerical and experimental results show excellent agreement. In particular, the clear attenuation of the wave field in the phononic region due to Bragg scattering can be observed in the case of a pulse centred within the BG, i.e. at $27.5$ kHz, confirming the presence of the numerically calculated BG. It is worth highlighting that as the excitation is provided at a single point, the waves impinge on the phononic region from a wide range of angles proving the ``complete'' nature of the BG.

\section{Conclusions}

In this paper, the existence of a complete BG for Lamb waves propagating in a phononic plate is provided by means of full wave field measurements. Extensive numerical simulations are presented to optimize the attenuating performance of a $12$-mm thick PVC plate with $4$ rows of unit cells of lattice parameter $a = 20$ mm, characterized by through-the-thickness cross-like cavities arranged in a square ring. Ultrasonic measurements on an ad-hoc machined phononic PVC plate are performed via SLDV measurements and compared to the numerical results. Excellent agreement between the numerical prediction and experimental measurements is found both in terms of the BG frequency range as well as the decreasing trend of the transmission power spectrum. Besides confirming the existence of the numerically predicted complete BGs, measurements provide a direct observation of the scattering phenomena observed for waves characterized by frequency content below, within and above the BG.

This design shows promise due to its simplicity and effectiveness, and could be exploited at different size scales, e.g. at large scales for seismic shielding applications. In \cite{Miniaci_NJP_2016} we have shown that for sandy type soils, similar structures in the range of about 10 m are sufficient for effective shielding at frequencies below 5 Hz. Experimental verification remains to be performed on scaled models of these shields. In future, it would be interesting to extend the experiments to other geometries of practical interest, and especially to weakly disordered phononic materials or stubbed phononic plates, in order to further extend the study of metamaterial geometries providing efficient attenuation over wide low-frequency ranges.


\vspace{1cm}

\textbf{Acknowledgements}

M.M. acknowledges funding from the European Union's Horizon 2020 research and innovation programme
under the Marie Skłodowska-Curie grant agreement no. 658483. N.M.P. is supported by the European Research Council PoC 2015 "Silkene" No. 693670, by the European Commission H2020 under the Graphene Flagship Core 1 No. 696656 (WP14 "Polymer composites") and FET Proactive "Neurofibres" grant No. 732344. F.B. is supported by H2020 FET Proactive "Neurofibres" grant No. 732344. 




\newpage

\begin{figure}[]
\centering
\subfigure
{\includegraphics[trim=28mm 105mm 60mm 18mm, clip=true, width=.9\textwidth]{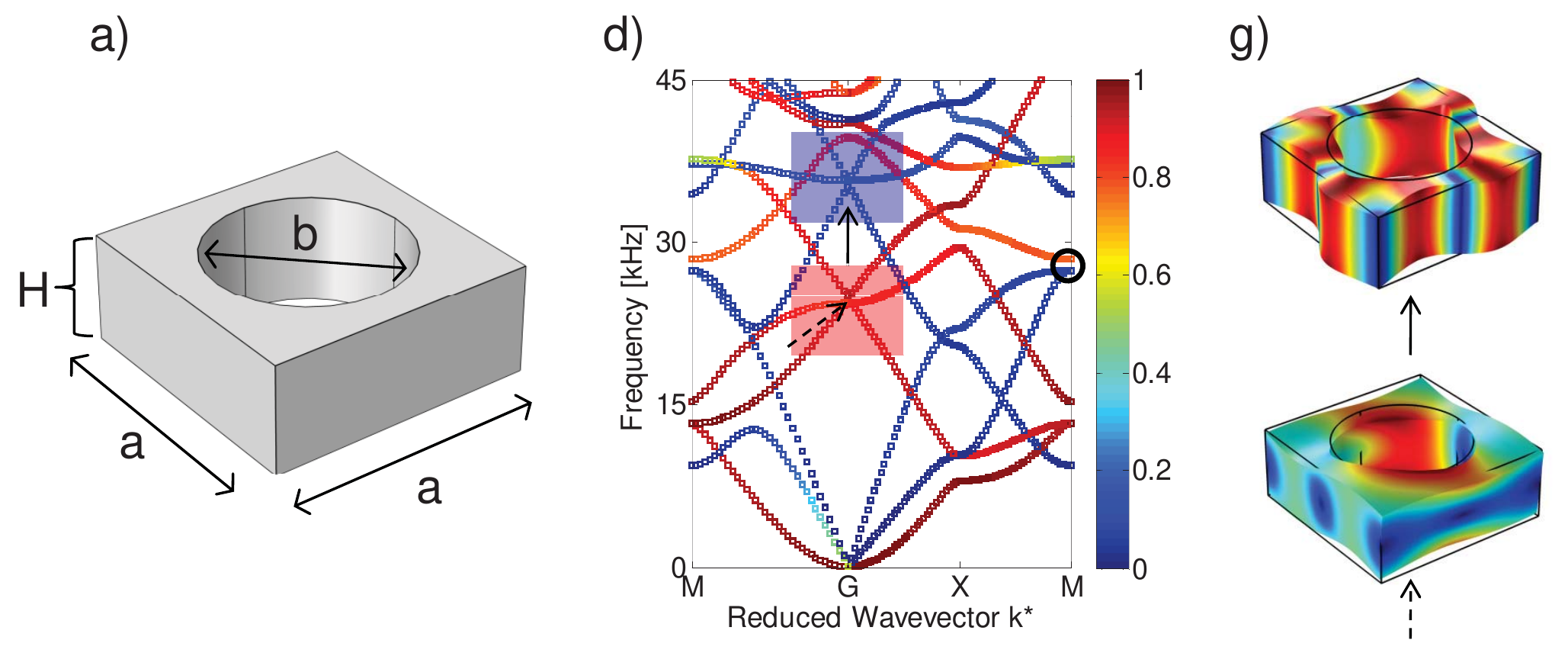}}
\subfigure
{\includegraphics[trim=28mm 105mm 60mm 18mm, clip=true, width=.9\textwidth]{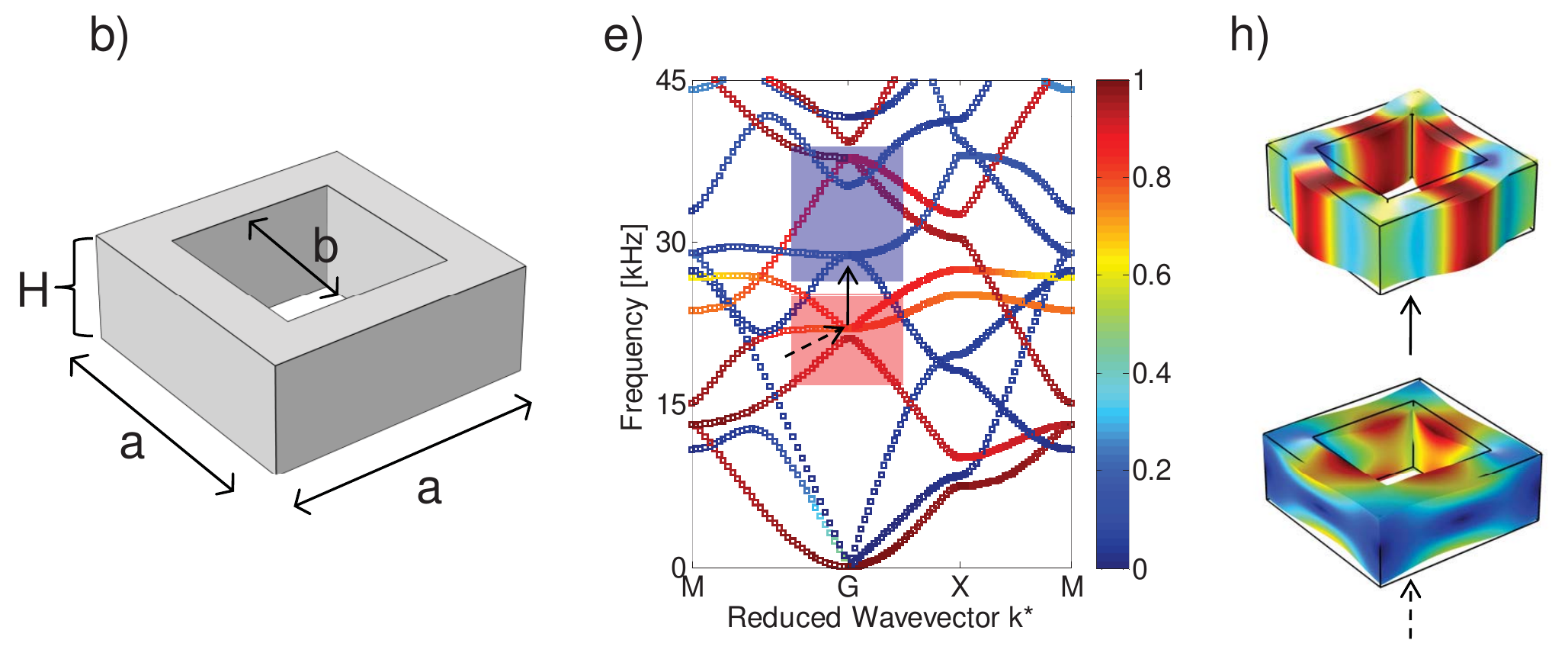}}
\subfigure
{\includegraphics[trim=28mm 105mm 60mm 18mm, clip=true, width=.9\textwidth]{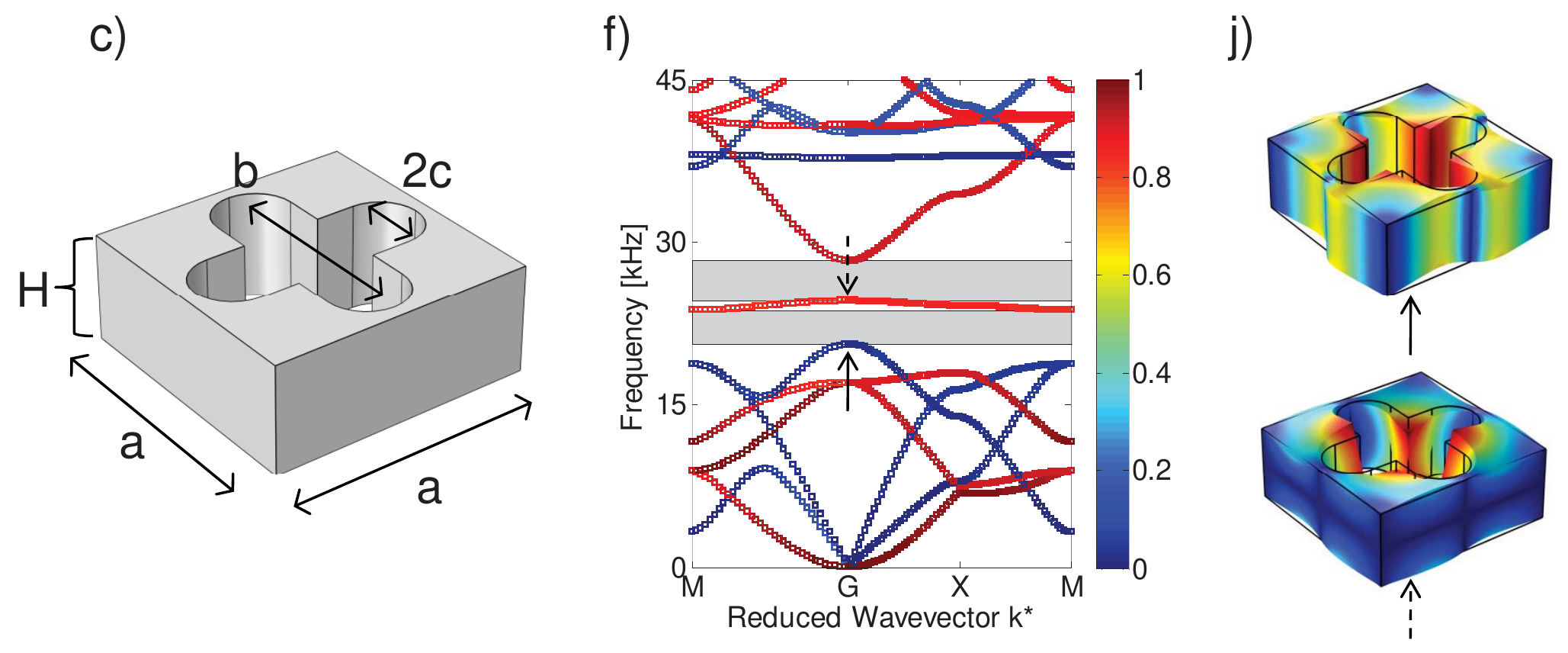}}
\caption{Unit cells made of (a) circular, (b) square and (c) cross-like cavities in a PVC matrix and their corresponding band structures (d-f) and mode shapes (g-j). The color of the curves indicates the mode polarization, ranging from pure in-plane (blue) to pure out-of-plane (red). On the other hand, blue and red in the mode shapes represent the zero and maximum displacement, respectively. Geometrical parameters are given in Table \ref{GeomPar} and guarantee the same filling fraction $f_f = 41\%$ for all the three models.}
\label{Fig1}
\end{figure}

\begin{figure}
\centering
\includegraphics[trim=0mm 0mm 0mm 0mm, width=0.5\textwidth]{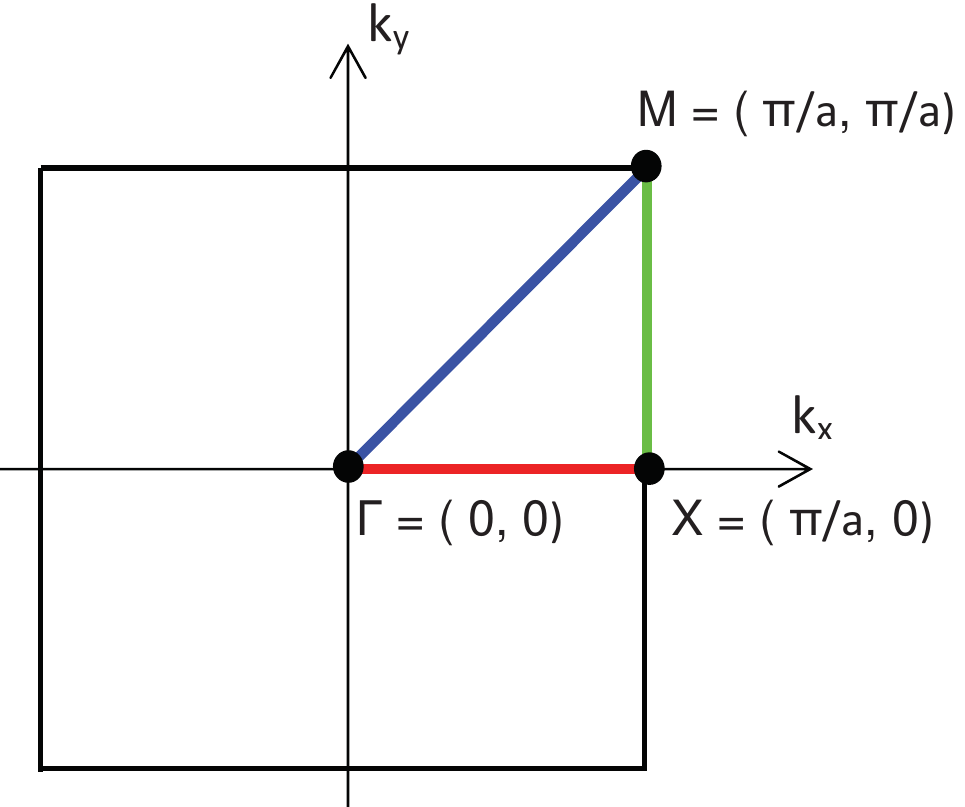}
\captionof{figure}{\label{FIBZ} Schematic representation of the first irreducible Brillouin zone for a square lattice.}
\end{figure}

\begin{figure}[]
\centering
\subfigure
{\includegraphics[trim=0mm 0mm 0mm 0mm, clip=true, width=.75\textwidth]{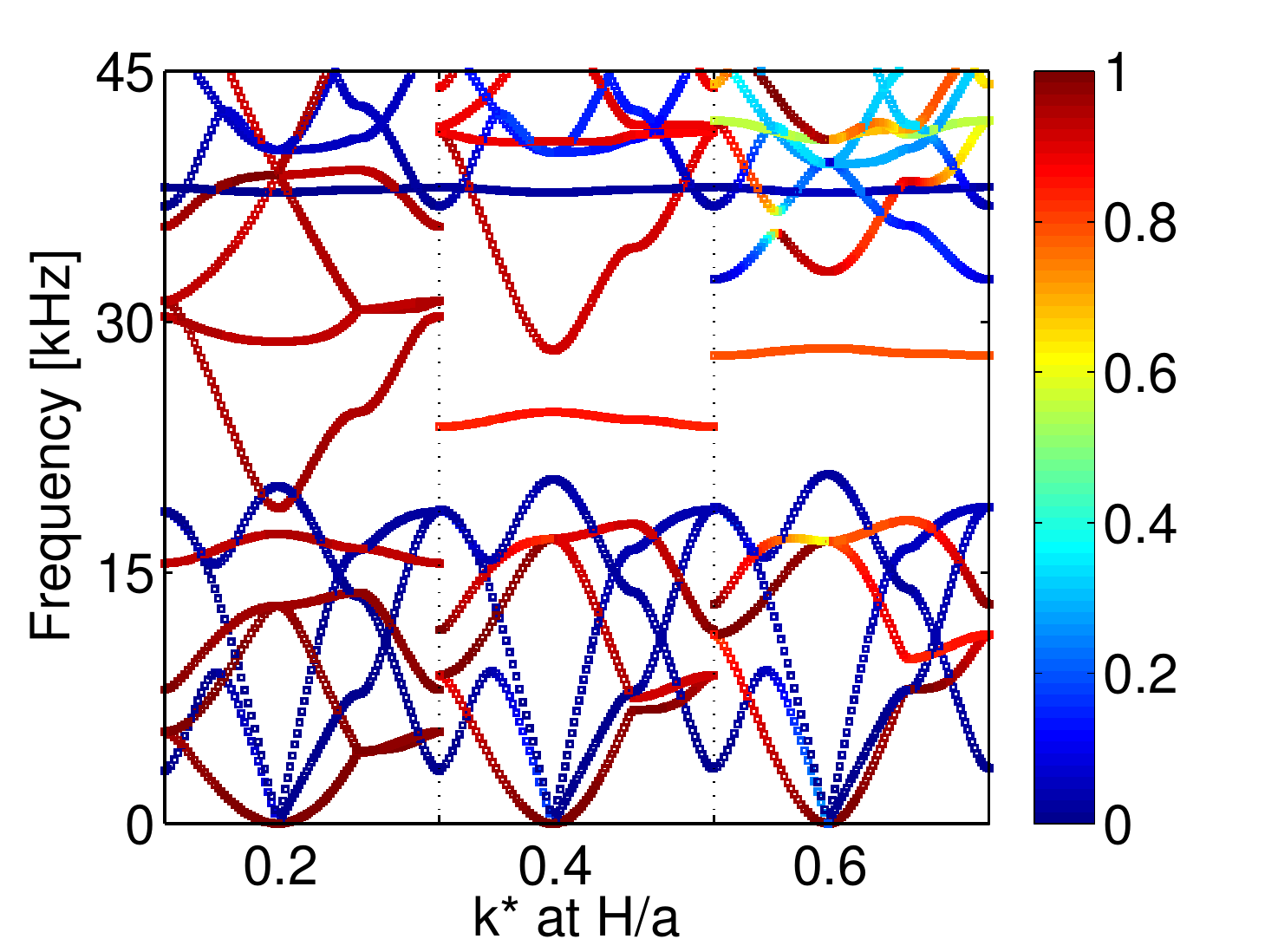}}
\put (-375,270) {\fontfamily{phv}\selectfont {\fontsize{16}{24}\selectfont a)}}
\\
\subfigure
{\includegraphics[trim=0mm 0mm 0mm 0mm, clip=true, width=.75\textwidth]{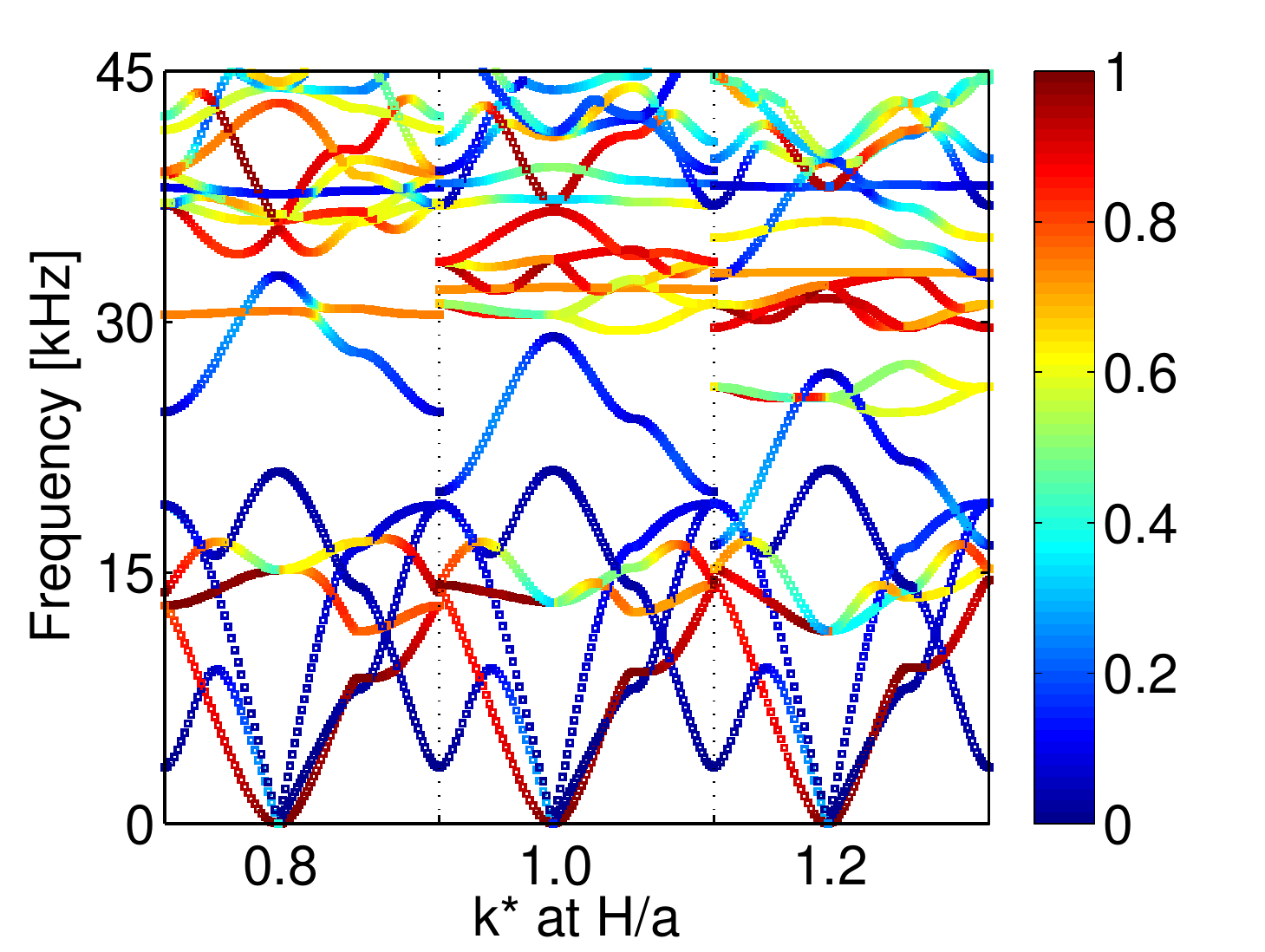}}
\put (-375,270) {\fontfamily{phv}\selectfont {\fontsize{16}{24}\selectfont b)}}
\caption{Parametric study showing the influence of the height to lattice parameter ratio $H/a$ on the band structure for the case study of a unit cell with a cross-like cavity. Dispersion diagrams for (a) $H/a = 0.2 - 0.6$ and (b) $H/a = 0.6 - 1.2$ are reported. The color of the curves indicates the mode polarization, ranging from pure in-plane (blue) to pure out-of-plane (red).}
\label{Fig3}
\end{figure}

\begin{figure}[]
\centering
{\includegraphics[trim=0cm 0cm 0cm 0cm, clip=true, width=.8\textwidth]{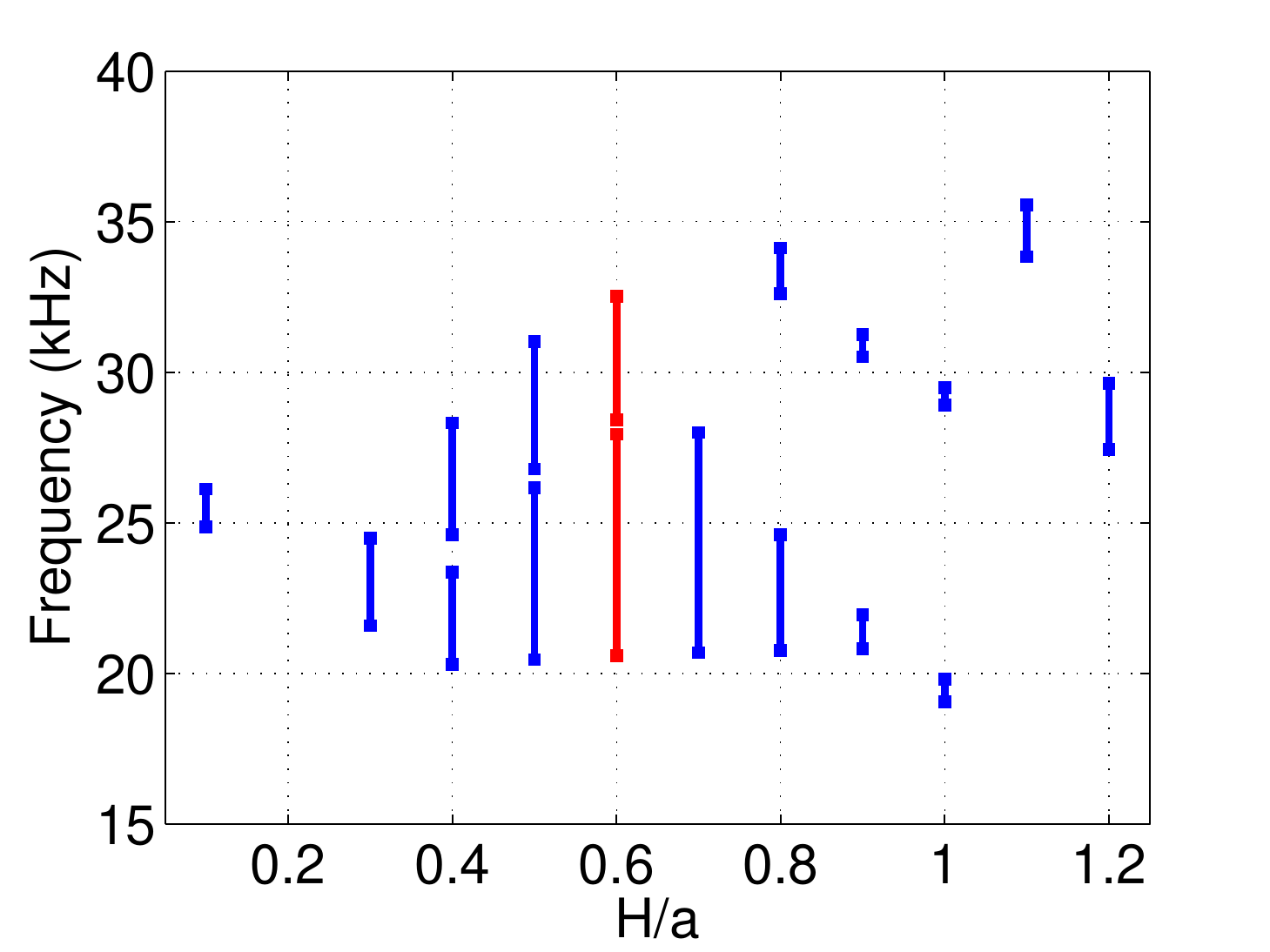}}
\caption{BG frequency versus the unit cell thickness to lattice parameter ratio $H/a$. The ratio leading to the widest total BG is highlighted in red.}
\label{WAZ_BGs}
\end{figure}

\begin{figure}[]
\centering
{\includegraphics[trim=14mm 105mm 15mm 11mm, clip=true, width=.9\textwidth]{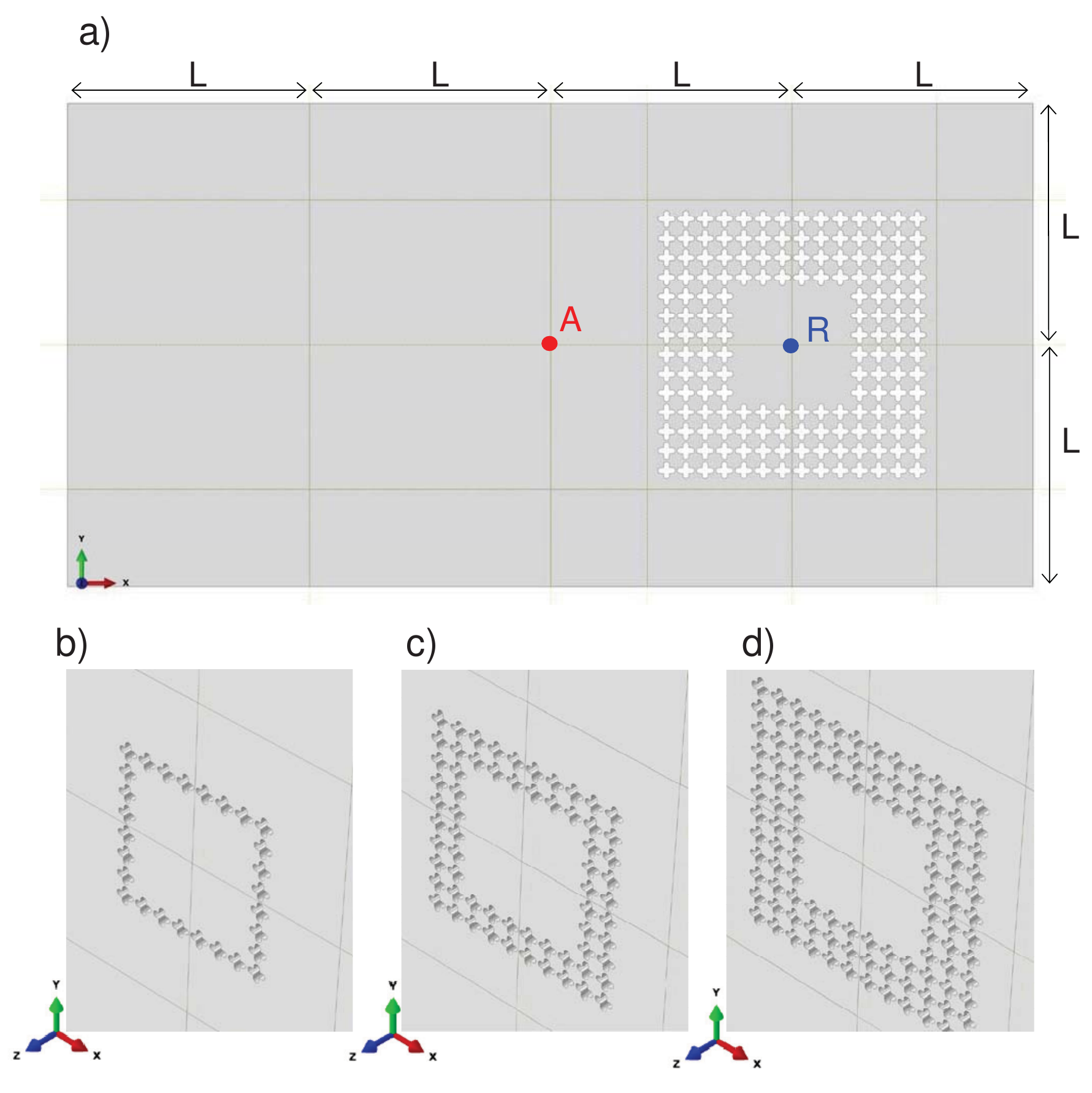}}
\caption{(a) Schematic representation of the numerical model of a plate including a 4-row "ring-like" area in the right part of the plate. A zoom of the same structure with the "ring-like" zone made of (b) $2$ rows, (c) $3$ rows and (d) $4$ rows of unit cells is provided. $A$ and $R$ denote the Lamb wave excitation and receiving points, respectively. $L = 250$ mm.}
\label{Rows_design}
\end{figure}

\begin{figure}[]
\centering
\subfigure
{\includegraphics[trim=0mm 0mm 0mm 0mm, clip=true, width=.48\textwidth]{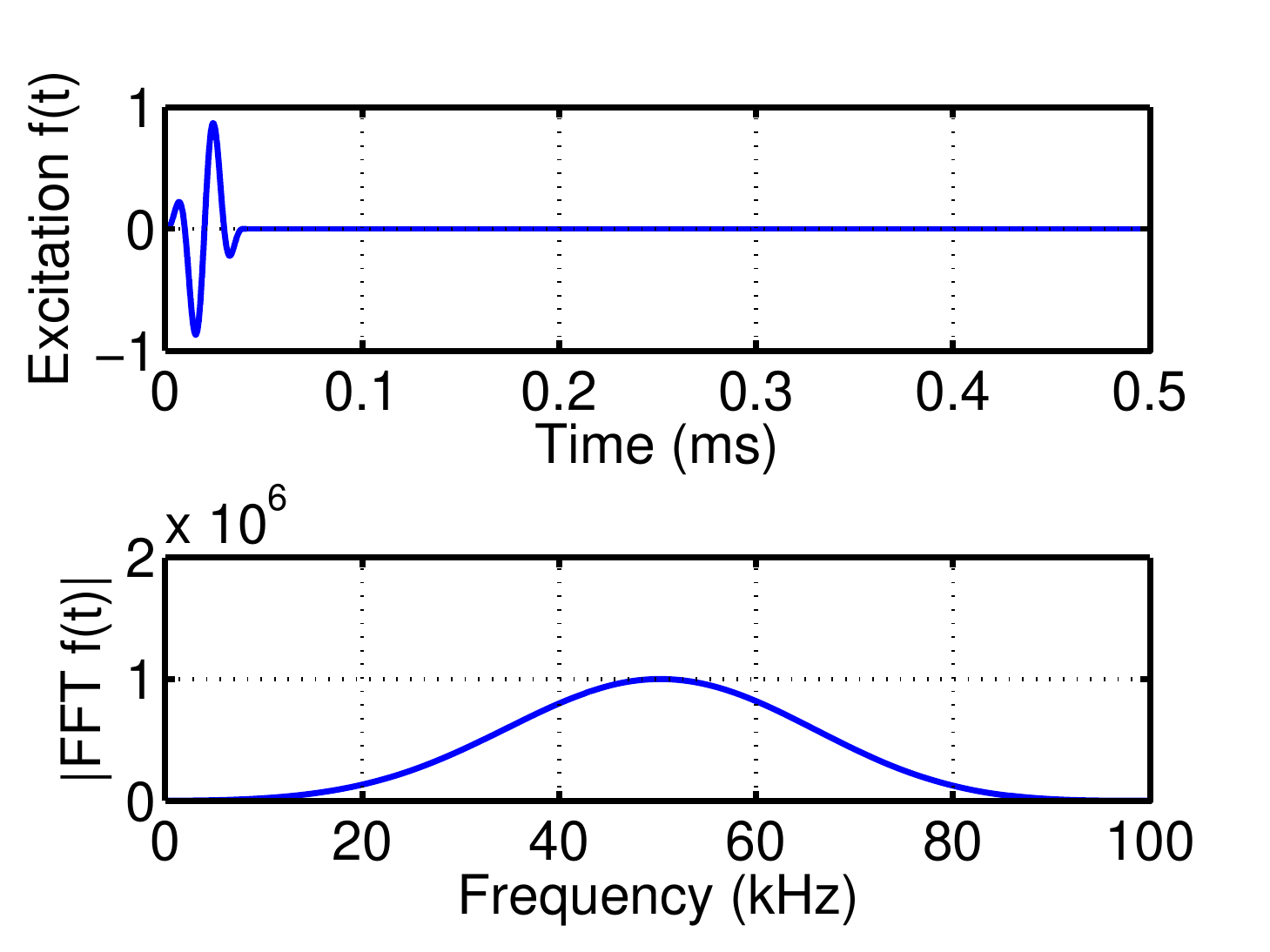}}
\put (-240,180) {\fontfamily{phv}\selectfont {\fontsize{13}{24}\selectfont a)}}
\subfigure
{\includegraphics[trim=0mm 0mm 0mm 0mm, clip=true, width=.48\textwidth]{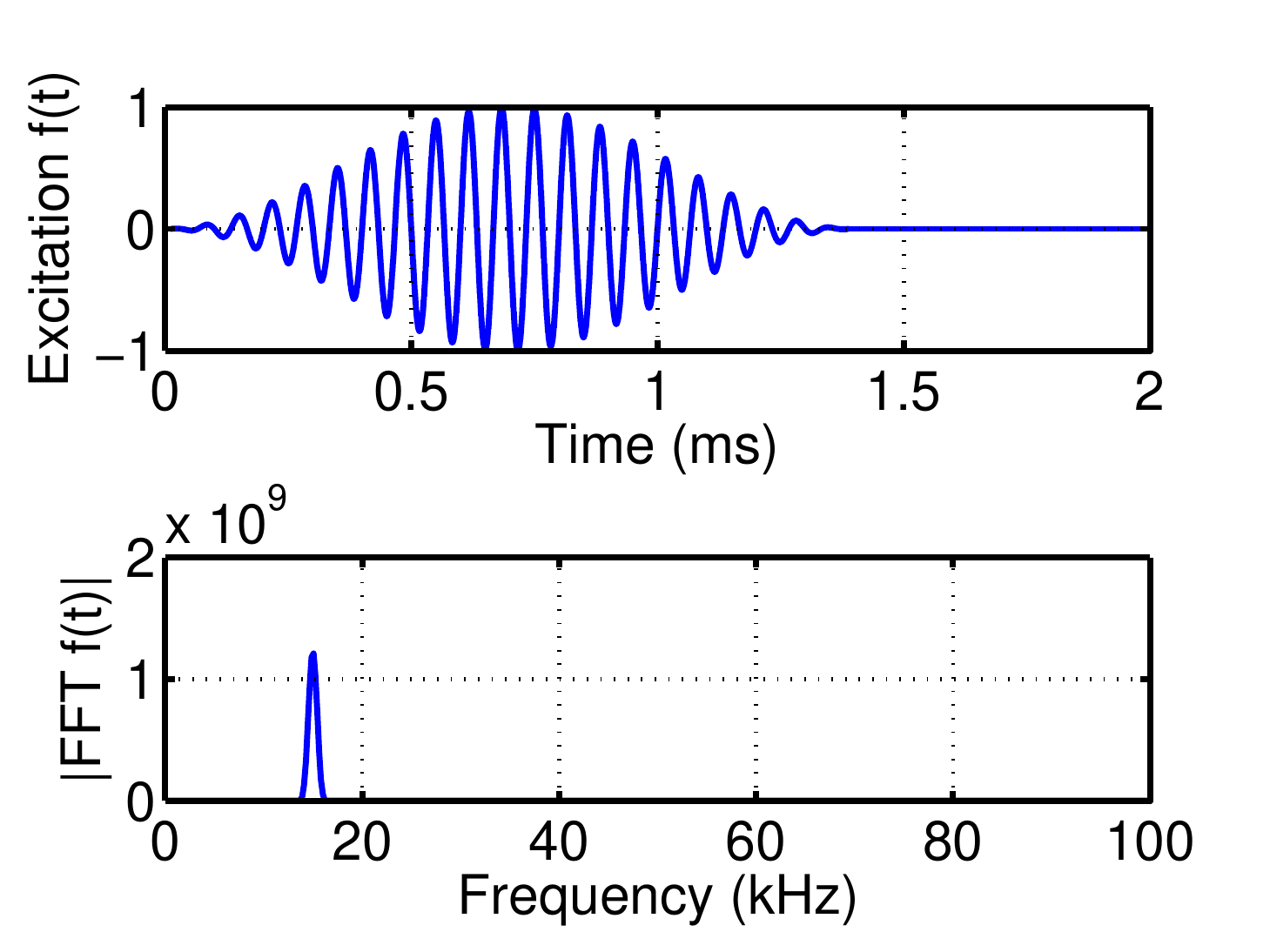}}
\put (-240,180) {\fontfamily{phv}\selectfont {\fontsize{13}{24}\selectfont c)}}
\\
\subfigure
{\includegraphics[trim=0mm 0mm 0mm 0mm, clip=true, width=.48\textwidth]{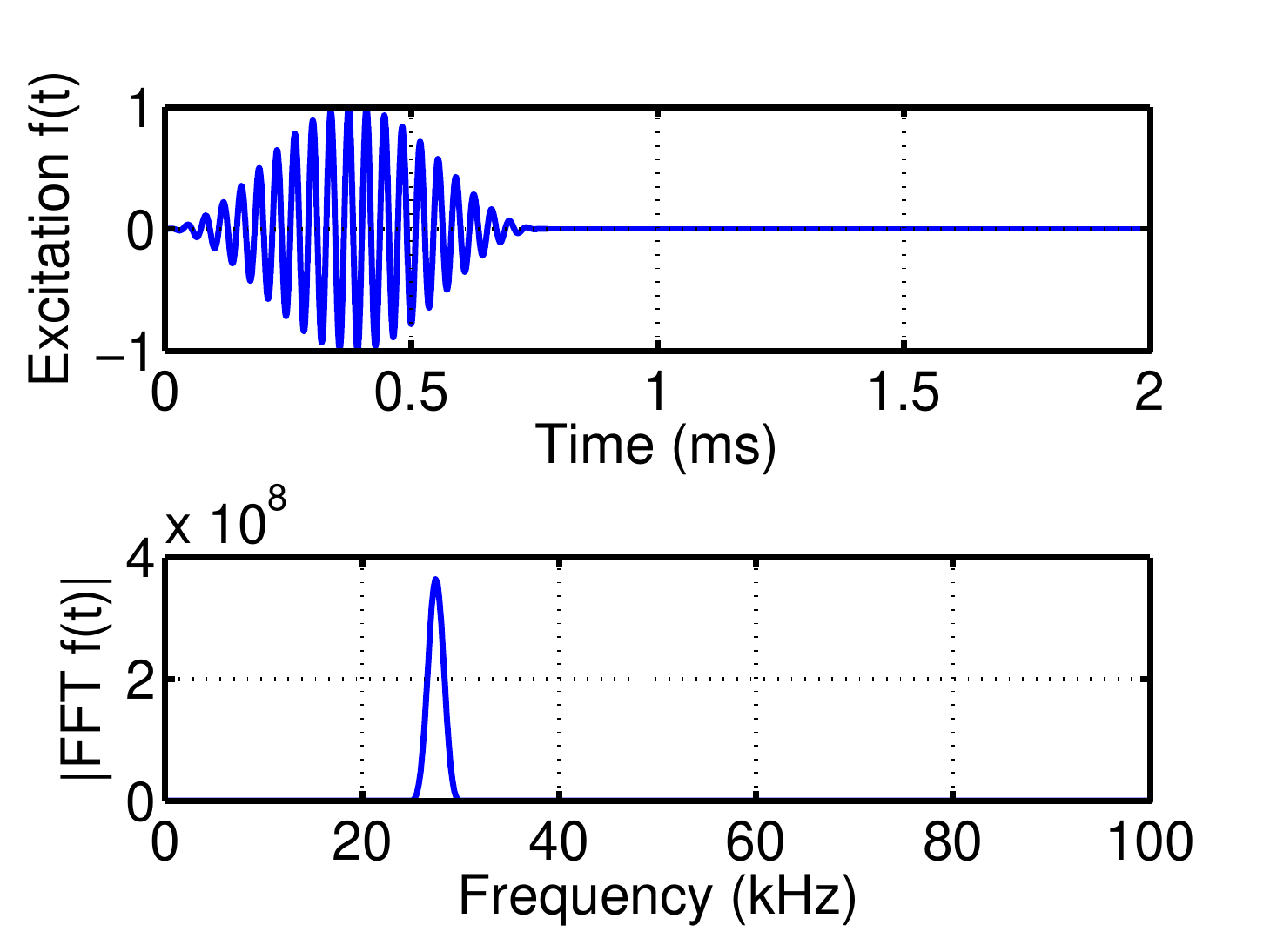}}
\put (-240,180) {\fontfamily{phv}\selectfont {\fontsize{13}{24}\selectfont b)}}
\subfigure
{\includegraphics[trim=0mm 0mm 0mm 0mm, clip=true, width=.48\textwidth]{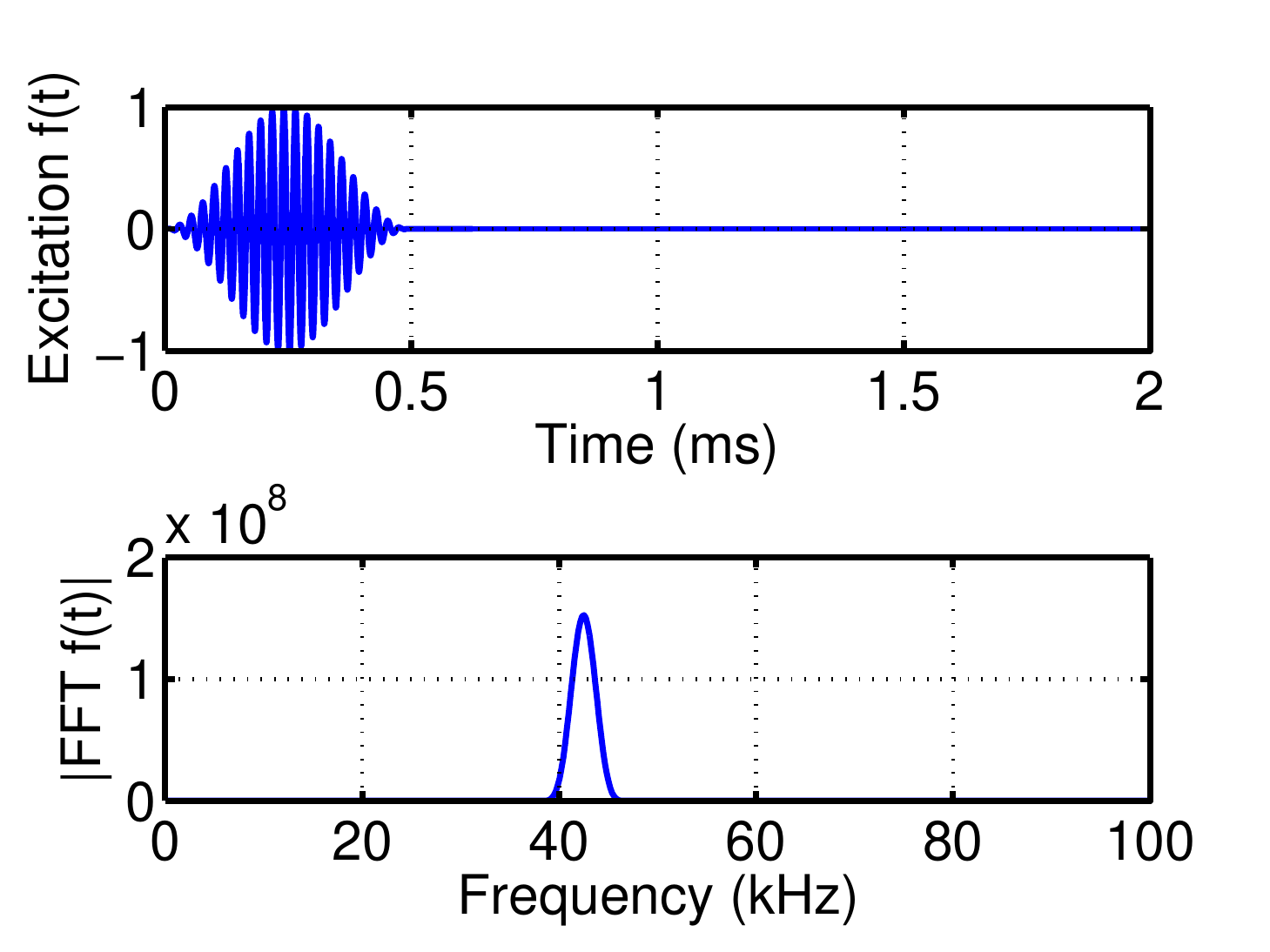}}
\put (-240,180) {\fontfamily{phv}\selectfont {\fontsize{13}{24}\selectfont d)}}
\caption{Excitation pulses: (a) $2$ sine cycles at $50$ kHz modulated by a Hanning window and $21$ sine cycles at (b) $27.5$ kHz (c) $15$ kHz and (d) $42.5$ kHz modulated by a Hanning window.}
\label{Pulses_PVC_plate}
\end{figure}

\begin{figure}[]
\centering
\subfigure
{\includegraphics[trim=0mm 0mm 0mm 0mm, clip=true, width=.75\textwidth]{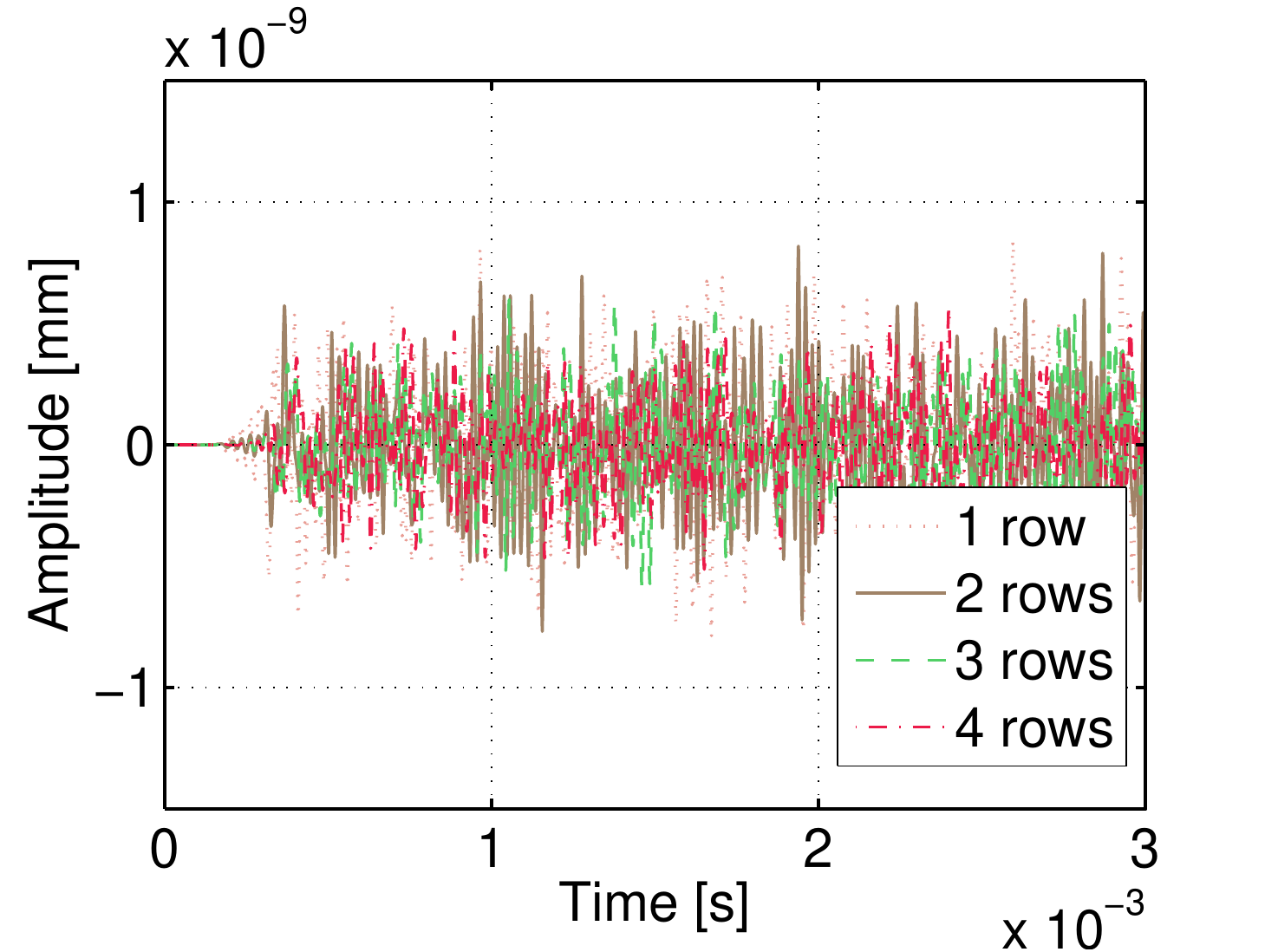}}
\put (-370,280) {\fontfamily{phv}\selectfont {\fontsize{16}{24}\selectfont a)}}
\\
\subfigure
{\includegraphics[trim=0mm 0mm 0mm 0mm, clip=true, width=.75\textwidth]{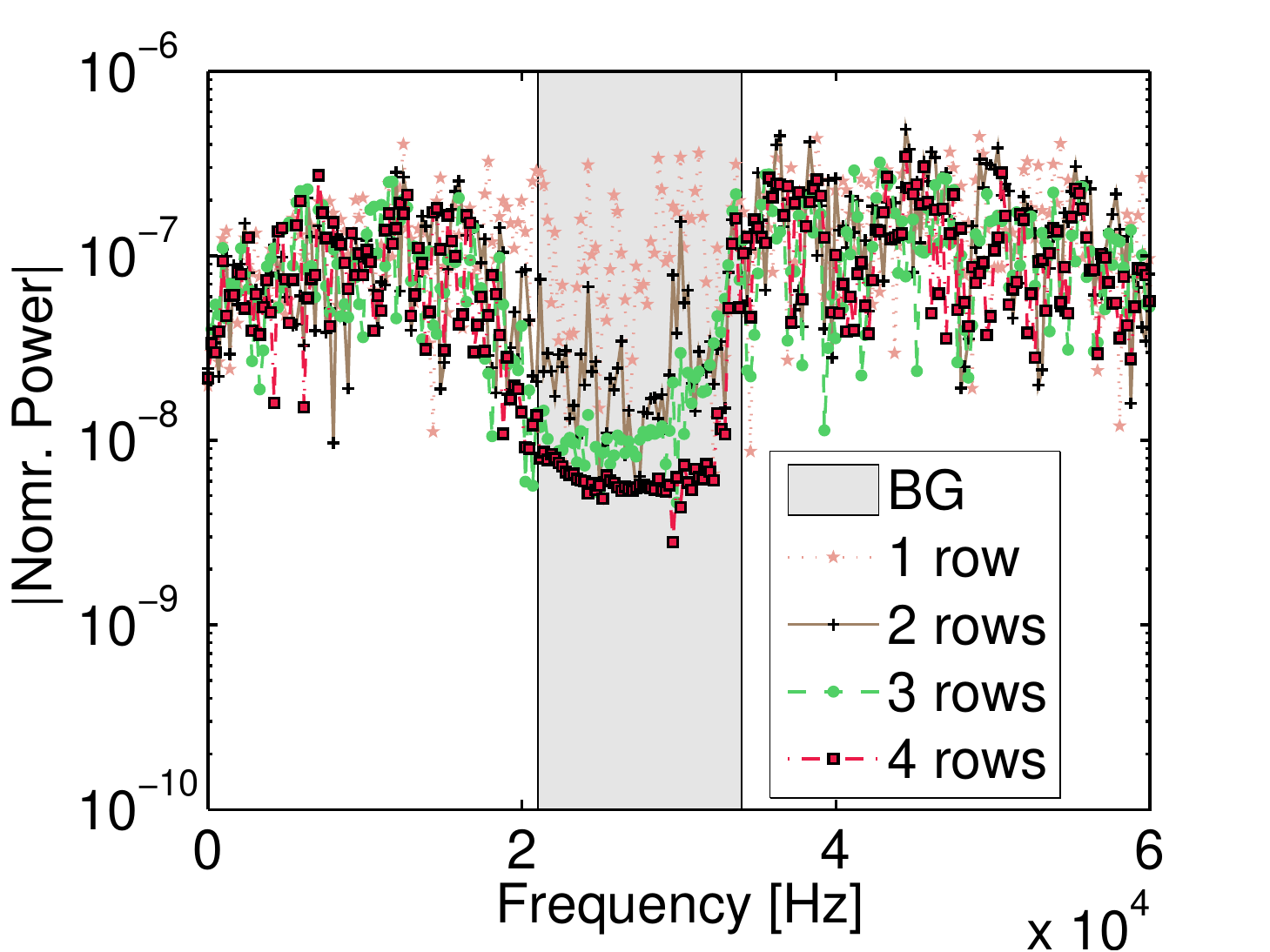}}
\put (-370,280) {\fontfamily{phv}\selectfont {\fontsize{16}{24}\selectfont b)}}
\caption{(a) Out-of-plane displacement for the phononic plate with $1, 2, 3$ and $4$ rows and (b) its Fourier Transform. The gray region denotes the calculated BG.}
\label{TimeHistory_FFT_Fc50kHz_Nc2}
\end{figure}

\begin{figure}[]
\centering
{\includegraphics[trim=20mm 51mm 23mm 20mm, clip=true, width=1\textwidth]{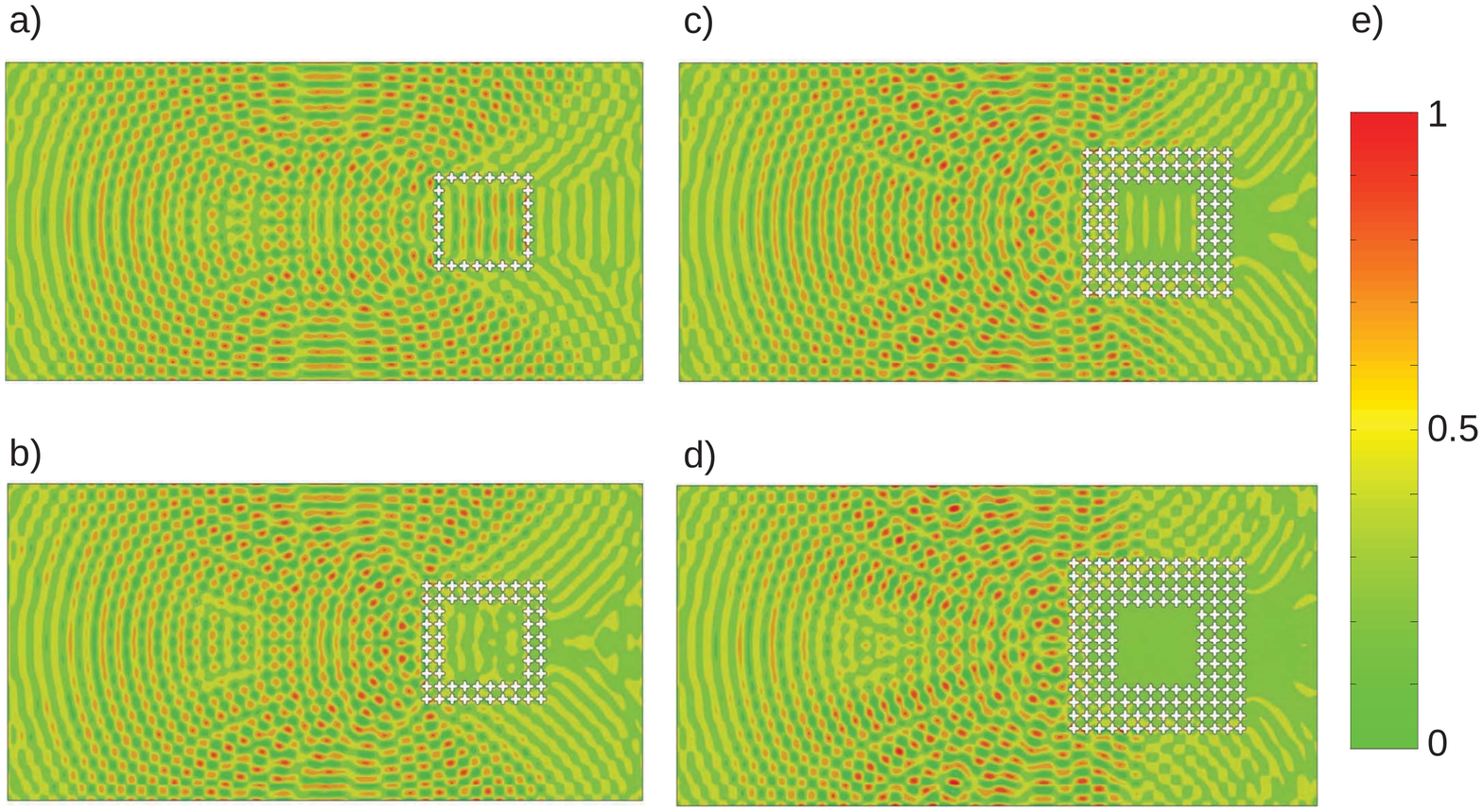}}
\caption{Von Mises stress for the phononic plate with (a) $1$ row, (b) $2$ rows, (c) $3$ rows and (d) $4$ rows. Displacement fields are recorded at $780$ $\mu s$ from the excitation time. (e) Colours from green to red denotes zero and maximum stress, respectively.}
\label{VonMises_stress_ParametricoFori}
\end{figure}

\begin{figure}[]
\centering
{\includegraphics[trim=22mm 110mm 72mm 20mm, clip=true, width=1\textwidth]{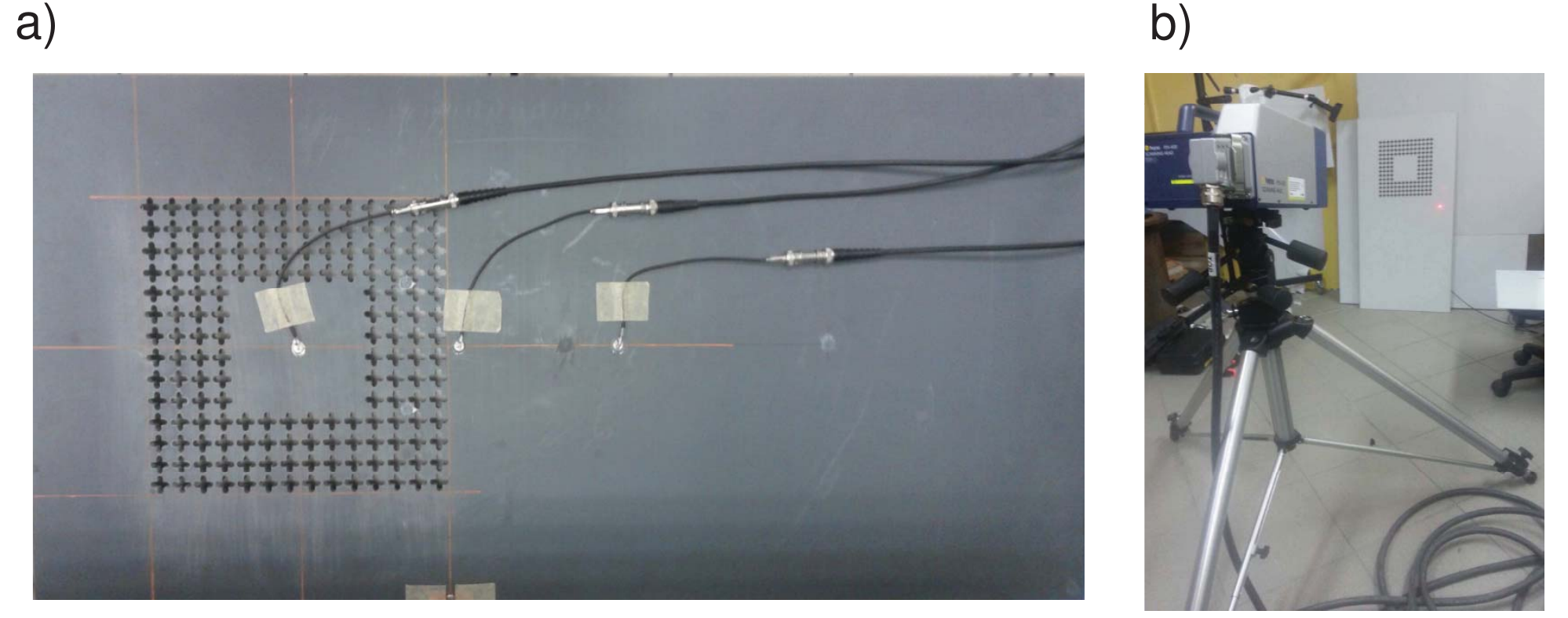}}
\caption{(a) In-plane view of the SLDV scanning grid point superimposed on the PVC phononic plate and (b) positioning of the SLDV measurement head.}
\label{piastra_PVC_reale}
\end{figure}

\begin{figure}[]
\centering
{\includegraphics[trim=30mm 42mm 46mm 18mm, clip=true, width=1\textwidth]{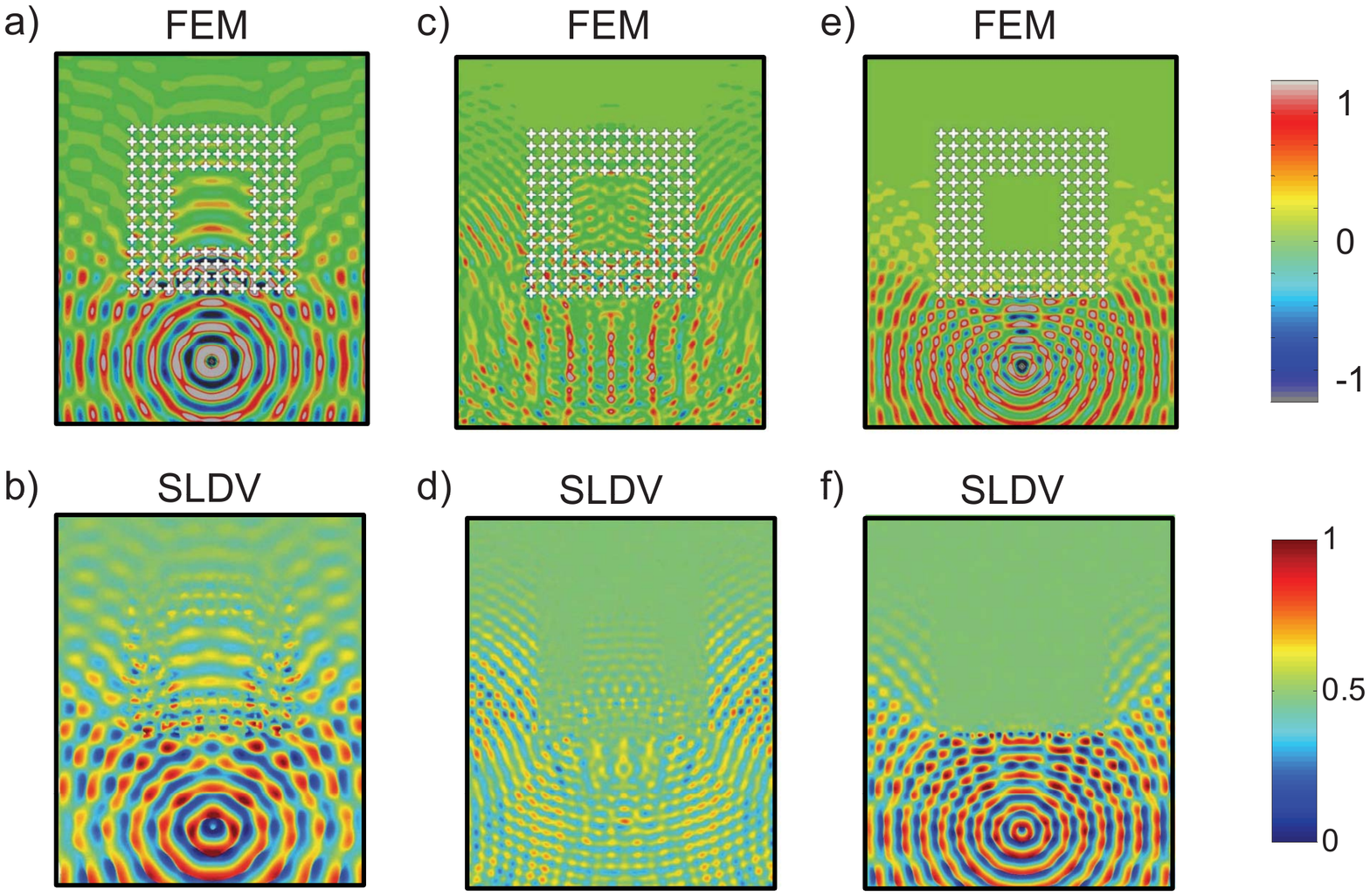}}
\caption{Computed and experimental out-of-plane displacement, respectively, for excitation below (a,b), inside (c,d) and above (e,f) the BGs. Numerical results are computed by means of ABAQUS and measurements are performed by means of a PSV 400 3D by Polytec SLDV.}
\label{Wavefield_Reconstruction_Cfr_Num_Exp}
\end{figure}

\centering
\begin{tabular}{c | c c c}
\hline
\multicolumn{1}{c|}{Geometrical} & \multicolumn{3}{c}{Length [mm]}\\ \cline{2-4}
					parameter    & $case (i)$       	& $case (ii)$   	& $case (iii)$		\\
\hline
a 								& 20				& 20				& 20				\\
h 								& 12				& 12				& 12				\\
b								& 14.48		 		& 12.83				& 18 				\\
c 								& /			 		& /	 				&  3		 		\\
\hline
\end{tabular}
\captionof{table}{\label{GeomPar} Geometrical parameters for unit cells shown in Figs. \ref{Fig1}a-c, characterized by (i) circular, (ii) square and (iii) cross-like cavities, respectively. The parameters are chosen so as to guarantee the same filling fraction for all the three designs.}

\end{document}